\documentclass[10pt, twocolumn]{LML}

\newcommand{\comment}[1]{}

\tikzset{
photon/.style={decorate, decoration={snake,amplitude=2pt, segment length=5pt}, draw=black},
particle/.style={draw=black, postaction={decorate}, decoration={markings,mark=at position .5 with {\arrow[draw=black]{>}}}},
antiparticle/.style={draw=black, postaction={decorate}, decoration={markings,mark=at position .5 with {\arrow[draw=black]{>}}}},
gluon/.style={decorate, draw=black, decoration={coil,amplitude=4pt, segment length=5pt}}
goldstone/.style={draw=green,postaction={decorate},decoration={markings,mark=at position .5 with {\arrow[draw=blue]{>}}}}
}

\begin{document}

\title{A fake doublet solution to the muon anomalous magnetic moment}

\author[a,b,c]{Damiano Anselmi\thanks{damiano.anselmi@unipi.it}}

\author[c]{Kristjan Kannike\thanks{kristjan.kannike@cern.ch}}

\author[c]{Carlo Marzo\thanks{carlo.marzo@kbfi.ee}}

\author[c]{Luca Marzola\thanks{luca.marzola@cern.ch}}

\author[c]{Aurora Melis\thanks{aurora.melis@kbfi.ee}}

\author[c]{Kristjan M\"u\"ursepp\thanks{kristjanmuursepp@gmail.com}}

\author[c]{Marco Piva\thanks{marco.piva@kbfi.ee}}

\author[c]{Martti Raidal\thanks{martti.raidal@cern.ch}}

\affil[a]{Dipartimento di Fisica ``E. Fermi'', Universit\`a di Pisa, Largo B. Pontecorvo 3, 56127 Pisa, Italy}
\affil[b]{INFN, Sezione di Pisa, Largo B. Pontecorvo 3, 56127 Pisa, Italy}
\affil[c]{Laboratory of High Energy and Computational Physics, NICPB, R\"avala 10, 10143 Tallinn, Estonia}

\date{\today}

\twocolumn[
\maketitle
\begin{onecolabstract}
Extensions to the Standard Model that use strictly off-shell degrees of freedom -- the fakeons -- allow for new measurable interactions at energy scales usually precluded by the constraints that target the on-shell propagation of new particles. Here we employ the interactions between a new fake scalar doublet and the muon to explain the recent Fermilab measurement of its anomalous magnetic moment. Remarkably, unlike in the case of usual particles, the experimental result can be matched for fakeon masses below the electroweak scale without contradicting the stringent precision data and collider bounds on new light degrees of freedom. Our analysis, therefore, demonstrates that the fakeon approach offers unexpected viable possibilities to model new physics naturally at low scales.
\end{onecolabstract}
]

\saythanks


\section*{Introduction} 
\label{sec:intro}
The longstanding anomaly concerning the magnetic moment of the muon, $a_\mu = (g-2)_\mu/2,$ is generally interpreted as the effect of new degrees of freedom that, coupling to the muon, leave their imprint in this precision observable. At the level of model building, the requirement of perturbative couplings forces the particles responsible for the signal to appear at energies not far above the electroweak scale. Consequently, the proposed $(g-2)_\mu$ explanations are often in tension with the null results of complementary collider and precision searches.

The presence of an anomaly in $(g-2)_\mu$ is supported by the final result of the Brookhaven E821 experiment~\cite{Bennett:2006fi},  $a_\mu^{\rm E821}= 116 592 089(63)\times10^{-11}$, which gives rise to a $3.7\sigma$ deviation $\Delta a_\mu^{\rm E821}=  a_\mu^{\rm E821} -  a_\mu^{\rm SM} = (279 \pm 76)\times10^{-11}$ when confronted with the worldwide consensus of the SM contribution~\cite{Aoyama:2020ynm}, $a_\mu^{\rm SM}= 116 591 810(43)\times10^{-11}$. The observation has been recently updated by the Muon $g-2$ experiment at Fermilab, which found~\cite{gm2-fermilab,Abi:2021gix}
\begin{equation}
\label{eq:newg-2}
    a_\mu^{\rm Fl}= 116592040(54)\times10^{-11} ,
\end{equation}
corresponding to a $3.3\sigma$ deviation from the SM prediction. The combined outcome of the two experiments~\cite{gm2-fermilab,Abi:2021gix}
\begin{equation}
\label{eq:comb}
    a_\mu^{\rm Fl+E821}= 116592061(41)\times10^{-11} ,
\end{equation}
leads to a $4.2\sigma$ discrepancy with the theory prediction:
\begin{equation}\label{eq:g-2Fl}
    \Delta a_\mu^{\rm Fl+E821}=  a_\mu^{\rm Fl+E821} -  a_\mu^{\rm SM} = (251 \pm 59)\times10^{-11}.
\end{equation}
This significant deviation could be explained by not yet understood low-energy hadronic physics~\cite{Borsanyi:2021}, or, as we propose, by new physics contributions.

To this purpose, we address the $(g-2)_\mu$ problem in a new framework that overcomes the limitations of conventional model building by relying on new, strictly-virtual, degrees of freedom: the {\it fakeons}. Fakeons were originally proposed to solve the problem of ghosts in renormalizable theories of gravity~\cite{Anselmi:2017ygm} and Lee-Wick theories~\cite{Anselmi:2018kgz,Anselmi:2017lia,Anselmi:2017yux}. Nevertheless, any new particle can be made a fakeon by adopting the required prescription for its propagator. 

Concretely, we consider an extension of the SM that contains a new fake scalar doublet. The latter does not acquire a vacuum expectation value (VEV) and, besides the gauge and Higgs bosons, couples significantly only to muons. Although the scalar sector of the theory matches that of the Fake Inert Doublet Model~\cite{Anselmi:2021} (fIDM), the resulting phenomenology strongly differs for the presence of a Yukawa coupling that breaks the $\mathbb{Z}_2$ symmetry of the fIDM and singles out the leptons of the second SM generation.  

As we show below, the fakeon doublet explains the $(g-2)_\mu$ measurement~\eqref{eq:newg-2} even in a mass range that, for usual particles, is precluded by the measurements of the $Z$- and $W$-boson decay widths. To demonstrate the case, we check our solution against the constraints that target deviations from lepton universality in the $Z$ and $W$ boson decays. In particular, we study the contributions of the fake doublet to $Z\to 2\mu,$ $Z\to 4\mu$ and to the $\tau$ and $\mu$ lepton decays, showing that these processes do not impose significant constraints on our result.

The present study could be extended to address the anomalies pertaining to the measurement of the muonic proton radius~\cite{doi:10.1146/annurev-nucl-102212-170627} and possible lepton universality violation in the decays of $B$ mesons~\cite{Aaij:2021vac}. It would be also of interest to consider a higher-order kinetic term for the fake doublet, with the purpose of solving also the SM hierarchy problem through the mechanism previously used in Lee-Wick extensions~\cite{Grinstein:2007mp,Carone:2008bs, Espinosa:2011js}. In fact, the fakeon prescription can also be used to consistently include fake ghost particles in the theory and prevent their on-shell propagation.   

\section*{A fake doublet extension of the Standard Model} 
\label{sec:fakeons}

We consider the following Lagrangian,

\begin{align}
\label{eq:lag0}
    \mathcal{L} =& \mathcal{L}_{\rm SM} + (D_\mu\Phi)^\dagger (D^\mu\Phi)-V - \left( y \,\bar{\ell}_{L}^\mu \, \Phi\, \mu_R + \text{H.c.}\right), \\
    \label{eq:V}
    V=& - m_1^2 \modu{H}^2 + m_2^2 \modu{\Phi}^2  + \lambda_1 \modu{H}^4 + \lambda_2 \modu{\Phi}^4 
    \notag
    \\
     +\lambda_3 & \modu{H}^2 \modu{\Phi}^2 + \lambda_4 \modu{H^\dagger \Phi}^2 +\frac{1}{2} \lambda_5 \left((H^\dagger \Phi)^2 + {\rm H.c.}\right),
\end{align}
where $\ell_L^\mu$ is the second-generation left-handed lepton doublet, $H$ is the SM Higgs doublet and $\Phi = (\phi^+, \phi^0)^T$ is a fakeon doublet which transforms as $\{1, 2, 1/2\}$ under the $SU(3)_c\times SU(2)_L\times U(1)_Y$ gauge group. We decompose the complex neutral component in its scalar and pseudoscalar parts,
\begin{equation}
   \phi^0 = \dfrac{\phi_H + i \phi_A}{\sqrt 2}.
\end{equation}
Since $\Phi$ acquires no VEV, the scalar fakeon masses are  
\begin{align}
\label{eq:m}
    m^2_i = & \, m_2^2 + \lambda_i v^2,\, i=\phi^{\pm},\, \phi_H,\, \phi_A, \\
    \lambda_{\phi^{\pm}} = &  \, \lambda_3 ,\notag \\
    \lambda_{\phi_H} = & \, \lambda_3+ \lambda_4 +\lambda_5 ,\notag\\
    \lambda_{\phi_A} = &  \, \lambda_3+ \lambda_4 -\lambda_5, \notag
\end{align}
where $v$ is the Higgs doublet VEV. We adjust $\lambda_2$ so as to ensure that the potential is bounded from below and implicitly set the values of $m_2$ and $\lambda_3$ by specifying $m_{\phi^{\pm}}$. The $\lambda_4$ and $\lambda_5$  couplings, which regulate the hierarchy of the fakeon masses, are assumed to vanish unless specified otherwise. More details on the scalar sector of the fIDM can be found in~\cite{Anselmi:2021}. 

To explain the anomaly, we couple $\Phi$ to the muon via a real Yukawa coupling $y$ and assume negligible couplings to the remaining SM fermions. The resulting new interactions 
\begin{align} \label{eq:lag1}
    \mathcal{L} \supset &
    -\left(
    \frac{y}{2} \, \bar{\nu}_\mu \, \mu \, \phi^+ 
    +
    \frac{y}{2} \, \bar{\nu}_\mu \, \gamma_5 \, \mu  \, \phi^+ 
    + \text{H.c.}
    \right)
    \\ \notag &
    -  
    \frac{y}{\sqrt{2}} \, \bar{\mu} \, \mu \, \phi_H
    - 
   i\frac{y}{\sqrt{2}} \, \bar{\mu}\,  \gamma_5 \, \mu \, \phi_A
\end{align}
suffice to fully capture the fakeon physics of $(g-2)_\mu$. 

\section*{Fakeons: main phenomenological features} 
\label{sec:pheno}

The fakeon propagator relies on a quantization prescription that differs from the Feynman recipe adopted for the SM fields. As a result, fakeons can mediate new interactions in the same way as usual particles do, but cannot be on-shell. Consequently, fakeons do not appear in initial and final states of physical processes, and, in particular, cannot leave any direct imprint in an experimental apparatus after their propagation. 

In particular, the fakeon doublet cannot directly contribute to the decay width of the $Z$ boson even for masses below $m_Z/2$: the decay $Z\to \Phi^\dagger\Phi$, allowed by gauge interactions, is forbidden by the fakeon quantization.  Therefore, the results of LEP experiments~\cite{Zyla:2020zbs} do not affect our solution, contrarily to the case of new ordinary particles. However, constraints on the properties of fakeons can still arise from their virtual contributions at collider and precision experiments. Besides the processes characteristic of fIDM~\cite{Anselmi:2021}, further phenomenological signatures arise from the breaking of the $\mathbb{Z}_2$ symmetry. For instance, the fakeon doublet mediates four-lepton final states decays of the $Z$ boson at the tree level, and  contributes to both the di-muon and invisible $Z$ decay widths radiatively. Additionally, loop diagrams that contain fakeons are also modified above every threshold associated with resonant contributions. In the case studied, the most evident impact is on the imaginary parts of the involved amplitude.

\section*{The anomalous magnetic moment of the muon} 
\label{sec:g-2}
\begin{figure}[t]
\centering
\includegraphics{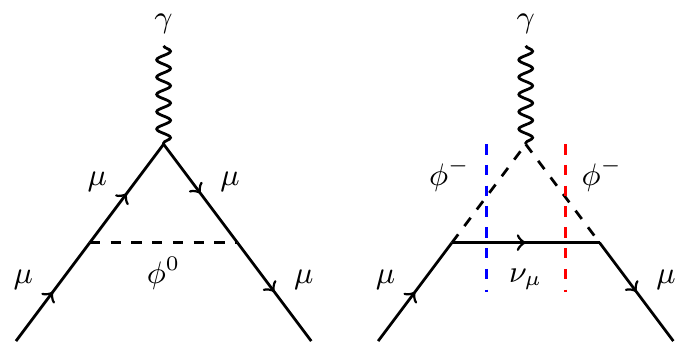}
\caption{Neutral and charged fake scalar contributions to the anomalous magnetic moment of the muon.
The dashed blue and red lines in the second diagram indicate the cuts corresponding to threshold contributions that vanish for the fakeon prescription.}
\label{fig:g-2}
\end{figure}

The interactions of the fakeon field in Eq.~\eqref{eq:lag1} affect the muon anomalous magnetic moment through the diagrams in Fig.~\ref{fig:g-2}, which account for the effect of the neutral and charged fake scalars, respectively. The total correction to the SM result, $\Delta a_{\mu}^\Phi=\Delta a_{\mu}^{H}+\Delta a_{\mu}^{A}+\Delta a_{\mu}^{\pm}$, is split into three terms.
The contributions of neutral components $\phi_H$ and $\phi_A$ are
\begin{align}\label{eq:Damu_H}\nonumber
\Delta a_{\mu}^{H}& =\frac{y^2}{16 \pi^2 r^2}\left[2 (r-1) (1-4r)h(1-4r)\right.\\
&\left.+r (3 r-2)+(3 r-1) \ln r\right], \quad r=\frac{m^2_{\mu}}{m_{\phi_H}^2},
\end{align}
\begin{align}\label{eq:Damu_A}\nonumber
\Delta a_{\mu}^{A}& =\frac{y^2}{16 \pi^2 r^2}\Big[2(3 r-1) h(1-4r)\\
&-r (2+r)+(r-1) \ln r\Big],\quad r=\frac{m^2_{\mu}}{m_{\phi_A}^2},
\end{align}
where $m_\mu\simeq 106$ MeV is the muon mass and $h(x)={\rm arctanh}(\sqrt{x})/\sqrt{x}$ for $0<x<1$,  $h(x)=\arctan(\sqrt{|x|})/\sqrt{|x|}$ for $x<0$. The relations above match the expressions in the literature for standard neutral scalar and pseudoscalar contributions -- see e.g.~\cite{Jegerlehner:2009ry,Lindner:2016bgg} -- because kinematics prevent this class of diagrams from developing an imaginary part. For the charged fakeon contribution, we have
\begin{equation}
\label{eq:Damu_cf}
    \Delta a_{\mu}^{\pm}=\frac{y^2}{32 \pi^2 r^2}\left[(r-2) r+2(r-1) \ln \modu{r-1}\right],
\end{equation}
where $r=m_{\mu}^2/m_{\phi^{\pm}}^2$. This relation {\it does} differ from the standard charged scalar contribution $\Delta a_{\mu}^{\pm, \,\mathrm{std}}$ on part of the parameter space by an imaginary part given by
\begin{equation}
\label{eq:Damu_Im}
    \Delta a_{\mu}^{\pm, \,\mathrm{std}}-\Delta a_{\mu}^{\pm}=-i\frac{ y^2}{16\pi}\frac{(r-1)}{r^2} \,\Theta(r-1),
\end{equation}
with $\Theta(x)$ being the Heaviside step function. In fact, for $m_{\phi^{\pm}} < m_\mu$ and vanishing neutrino masses, the process generally allows an imaginary part associated with the cuts shown in the second diagram of Fig.~\ref{fig:g-2}. For a standard scalar field this corresponds to the $\mu\to\phi \nu$ decay. The fakeon prescription forces these imaginary parts to vanish, so that $\Delta a_{\mu}^{\pm}$ is necessarily real. For usual particles, instead, the same cuts yield a finite imaginary contribution that redefines the form factor actually probed by the experiments~\cite{Avdeev:1998sx}. Such imaginary contributions would furthermore alter the muon lifetime in an external magnetic field~\cite{Binosi:2007ye}.

\begin{figure}[t]
    \centering 
    \includegraphics[scale=0.47]{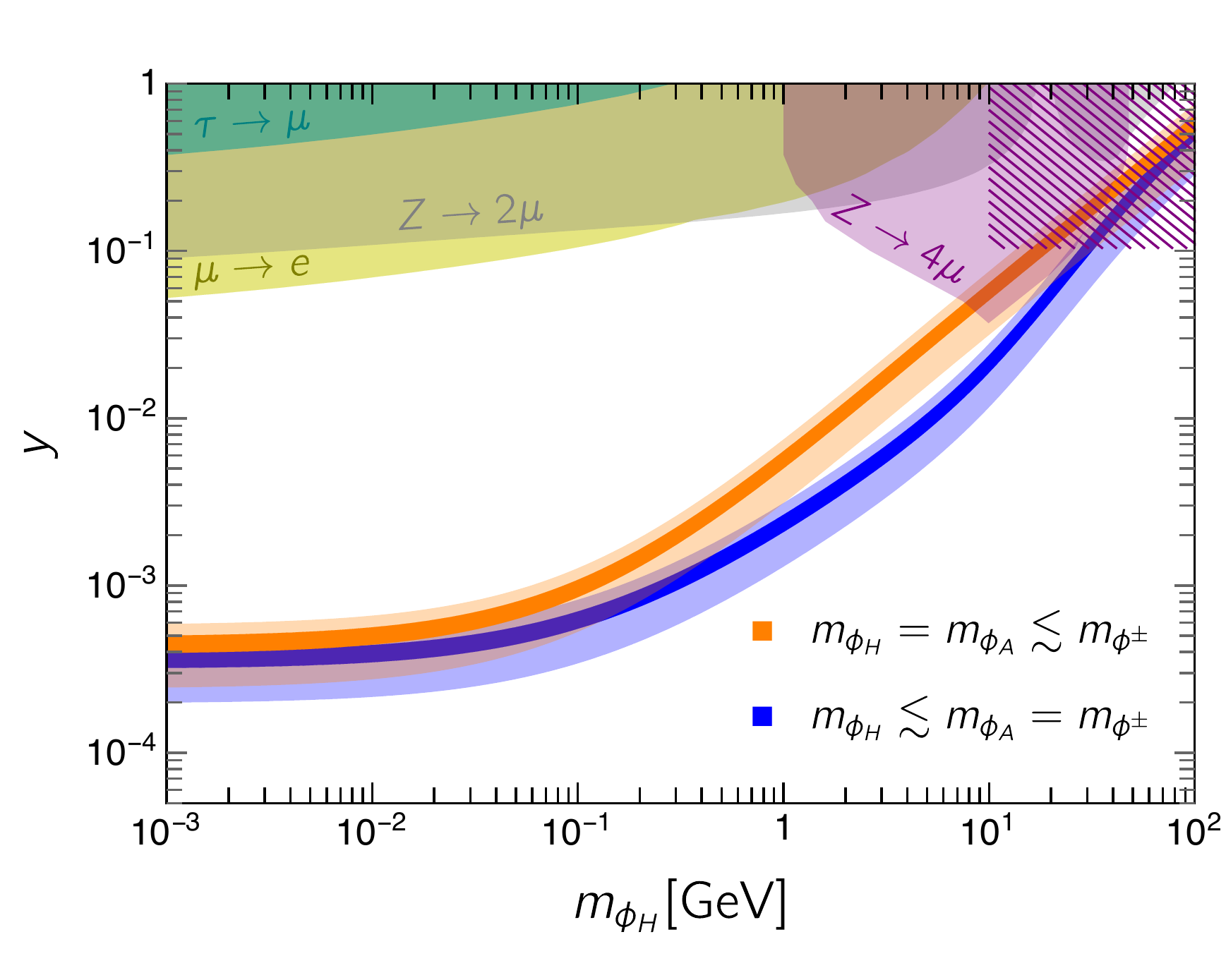}
    \caption{Regions of the parameter space $(m_{\phi_H},y)$ where the fake doublet contribution to $a_\mu$ falls within the $1\sigma$  and $3\sigma$ bounds on $\Delta a_{\mu}^{\rm Fl}$ for two representative  cases: $\lambda_4=-0.0003$ and $\lambda_5=0$ (orange) and $\lambda_4=\lambda_5=-0.002$ (blue). The remaining shaded regions represent the bounds from the indicated measurements.}
    \label{fig:g-2_y_vs_mphi} 
\end{figure}

In Fig.~\ref{fig:g-2_y_vs_mphi}, we show the parameter space where the total fakeon contribution $\Delta a^{\Phi}_\mu$ enters the $1\sigma$ and $3\sigma$ intervals allowed by Eq.~\eqref{eq:g-2Fl}, for two benchmark fakeon mass spectra. The orange band corresponds to a configuration where  $m_{\phi^{\pm}}\gtrsim m_{\phi_A}=m_{\phi_H}$, obtained by setting the quartic couplings as specified in the figure caption. The choice ensures that $m_{\phi^\pm}\geq 3$ GeV, as required by electroweak precision tests~\cite{Anselmi:2021}. The blue region shows the case of a different splitting, $m_{\phi_H}\lesssim m_{\phi_A}=m_{\phi^\pm}$, obtained for different values of the quartic coupling that result in $m_{\phi^\pm}\geq 10$ GeV. The remaining shaded areas denote the exclusion bounds due to the experimental searches discussed below. The results obtained for Yukawa couplings $y\lesssim 0.1$, corresponding to (CP-even) neutral fakeon masses below $\sim10$ GeV are mostly constrained only by the mentioned electroweak precision tests, which force a splitting of the charged fakeon component but do not bound the masses of the neutral ones~\cite{Anselmi:2021}. In both the analyzed cases, the observed values of $(g-2)_\mu$ are matched through the dominant contribution of the neutral CP-even fake component. 

We remark that our solutions are stable under radiative corrections because the negligible Yukawa couplings assumed for heavier SM fermions, as well as the absence of a VEV for the fake doublet, preclude new large two-loop Barr-Zee type contributions~\cite{Ilisie:2015tra}.

\section*{Collider and lepton flavor universality constraints} 
\label{sec:collider}

As previously emphasized, the decays $Z\to\phi_H\phi_A$ and $Z\to\phi^+\phi^-$, allowed by gauge interactions, cannot occur. Therefore, the $Z$-boson decays~\cite{Zyla:2020zbs} can only constrain the fakeon properties through their {\it virtual effects} in tree-level or loop processes yielding, in our case, muon final states. In the following we analyze the most important examples of these contributions.

\begin{figure}[t]
\centering
\includegraphics{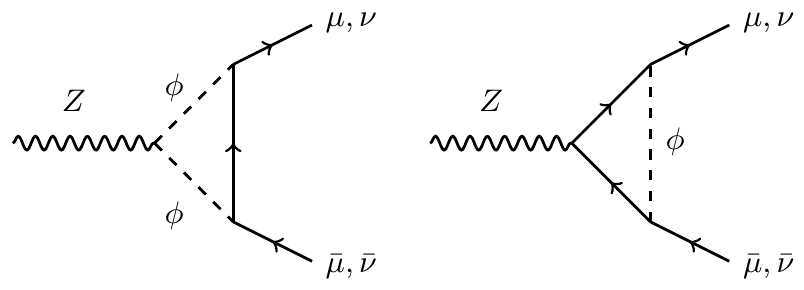}
\caption{The fake scalar doublet contributions to the decays of the $Z$-boson into $2\mu$ and $2\nu$.}
\label{fig:Z-2l} 
\end{figure}

The LEP measurements of the $Z$-boson leptonic decay widths provide stringent constraints on departures from lepton flavor universality. In the model at hand, muon final states receive new loop-level contributions from the diagrams depicted in  Fig.~\ref{fig:Z-2l}, which potentially unbalance the yield of the corresponding $Z$-boson decay channels. The corresponding departure from lepton universality is constrained by~\cite{Zyla:2020zbs}
\begin{equation}
    \label{LEP}
    R_{\mu e}=\frac{\Gamma(Z\rightarrow\mu^+\mu^-)}{\Gamma(Z\rightarrow e^+e^-)}=1.0001\pm 0.0024, 
\end{equation}
for the di-muon final state. Enhancements in the di-neutrino final state are, instead, constrained by the $Z$-boson invisible decay width $\Gamma({\rm inv})=499.0\pm 1.5$~MeV.

In order to assess the bound~\eqref{LEP}, we have calculated the leading order deviations of $R_{\mu e}$ from the SM value. The fakeon contribution sourced by the diagrams in Fig.~\ref{fig:Z-2l} is quantified in~\cite{Abe:2015oca}
\begin{equation}
\label{eq:DRmue}
    \Delta R_{\mu e}= \frac{g^2_Z \, m_Z}{6\pi}\frac{c_V  {\rm Re}[c^\Phi_V]+ c_A{\rm Re}[c^\Phi_A]}{\Gamma(Z\rightarrow\mu^+\mu^-)_{\rm SM}},
\end{equation}
where $g_Z$ is the neutral current coupling constants, $c_V=2\,s^2_W-1/2$, $c_A=-1/2$ and $c^\Phi_{V,A}$ are, respectively, the coefficients of the terms proportional to $\gamma_\mu$ and $\gamma_\mu \gamma^5$ in the obtained one-loop amplitude. The kinematics shows that the three internal lines of the diagrams cannot be all simultaneously on-shell. Therefore, the fakeon prescription affects only the imaginary part of the amplitude. Because Eq.~\eqref{eq:DRmue} is sensitive only to the real contribution, we can employ the relation to constrain the properties of the model by requiring that $\modu{\Delta R_{\mu e}}<0.0012$. The corresponding bound is indicated by the gray region in Fig.~\ref{fig:g-2_y_vs_mphi}. As we can see, the limit is far from excluding our solutions to $(g-2)_\mu$. Similarly, the new contribution to the $Z$-boson invisible width is also not constraining. We remark that for the case of usual scalar fields, these solutions would be precluded by the modifications to the tree-level $Z$-boson decay width resulting from the direct production of the new scalars. 

\begin{figure}[t]
\centering
\includegraphics{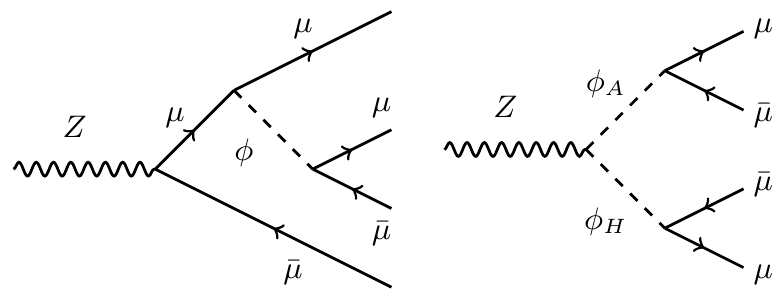}
\caption{$Z$-boson decays to $4\mu$ mediated by the fake scalar doublet.}
\label{fig:Z-4l}
\end{figure}

We analyze next the impact of the new fake doublet on decays of the $Z$-boson yielding $4\mu$ final states, which receive the new contributions shown in Fig.~\ref{fig:Z-4l}. The most precise measurements of the $Z\to 4\mu$ decay width from the LHC experiments~\cite{CMS:2012bw,Khachatryan:2016txa,Sirunyan:2017zjc,Aad:2014wra} indicate a branching ratio $BR(Z\to 4\mu)= (4.58\pm 0.26) \cdot 10^{-6}$, whereas the SM prediction is $BR(Z\to 4\mu)= (4.70\pm 0.03) \cdot 10^{-6}$~\cite{Rainbolt:2018axw}. 

The diagrams of Fig.~\ref{fig:Z-4l} must be summed, then multiplied by the complex conjugates and integrated over the phase space of the four muons. The result is a sum of cut diagrams with fakeons circulating in loops, which must be evaluated with the fakeon prescription. The usual techniques are not immediately applicable, but the impact of the result on our analysis can be estimated by means of simpler arguments. On dimensional grounds, we have contributions of the form
\begin{equation}
\label{eq:DAestimate}
\Delta BR(Z\to 4\mu)\lesssim\frac{m_Zy^4{\tilde g}^2}{(4\pi)^3\Gamma_Z},
\end{equation}
where ${\tilde g}$ is the coupling in the gauge vertex. For $y\sim 0.1$, we obtain a correction to $BR(Z\to 4\mu)$ of order $10^{-6}$, which is comparable to the experimental precision. Conservatively, we infer that values smaller than $y\sim 0.1$ are compatible with the constraint. We highlight this bound in Fig.~\ref{fig:g-2_y_vs_mphi} with a meshed region.

The same process was also recently studied in Ref.~\cite{Rainbolt:2018axw}, which quantified the contribution of the first diagram in Fig.~\ref{fig:Z-4l} in a toy model where the fake doublet is replaced by a usual scalar. The result is shown in Fig.~\ref{fig:g-2_y_vs_mphi} in purple. Due to the specific kinematics of the process, and since the CMS and ATLAS experiments impose a cut on the invariant mass of lepton pairs of $m_{\mu\mu}<4$~GeV and  $m_{\mu\mu}<5$~GeV, respectively, the constraint peaks in a narrow window around 10~GeV. Regardless of the details pertaining to the treatment of the new di-muon resonance in their simulation, we expect the bound to hold, at least, at the order of magnitude level.   

\begin{figure}[t]
\centering
\includegraphics{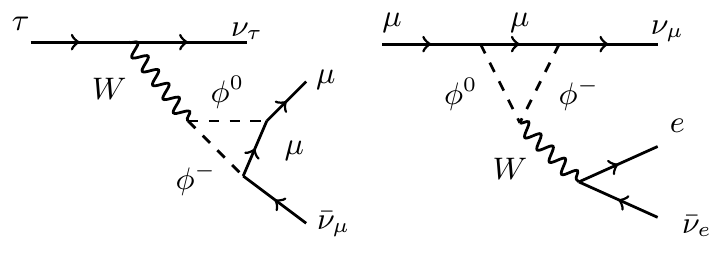}
\caption{ Fake doublet contributions to $\tau$ and $\mu$ decay.}
\label{fig:w-numu}
\end{figure}

The last precision observables that we analyze probe deviations from lepton flavor universality in the decays of the $\tau$ and $\mu$ leptons. The relevance of these bounds follows from the modifications in the muon sector due to the diagrams in Fig.~\ref{fig:w-numu}. Since no thresholds are involved in the limit of vanishing $W$ squared momentum in which the observables are matched, the relevant loop contributions are not modified by the fakeon prescription. 

To quantify the constraints, we compute the Wilson coefficients of the effective four-fermion interactions for the two processes. The dimension-$6$ operator for the decay of a charged lepton $\ell=\tau (\mu)$ into a lepton $\ell'=\mu(e)$ is~\cite{Pich:2013lsa}
\begin{align}
\label{eq:dim6op}
    \mathcal{L}_{\ell\to \ell^{\prime}}=4 \frac{G_{\ell^{\prime}\ell}}{\sqrt{2}}\Big[ & g_{RR}^{S,\ell^{\prime}\ell}\left(\overline{\ell^{\prime}_R}\nu_{\ell^{\prime} L}\right)\left(\overline{\nu_{\ell L}}\ell_{R}\right)\\ \notag
  & +g_{LL}^{V,\ell^{\prime}\ell}\left(\overline{\ell^{\prime}_L}\gamma^{\sigma}\nu_{\ell^{\prime} L}\right)\left(\overline{\nu_{\ell L}}\gamma_{\sigma}\ell_{L}\right)\Big],
\end{align}
where $G_{\ell^{\prime}\ell}$ is the effective Fermi constant and $g_{RR}^{S,\ell^{\prime}\ell}$ and $g_{LL}^{V,\ell^{\prime}\ell}$ are the Wilson coefficients due to SM and fakeon contributions. We compare our results to the experiments by using the Michel parameters \cite{Michel:1949qe,Kinoshita:1957zz,Kinoshita:1957zza}. The most constraining in our case is $\rho$, which, for the operators in \eqref{eq:dim6op}, reduces to
\begin{equation}
    \rho_{\ell\to\ell^{\prime}}=\frac{3}{4}\left(\frac{1}{4}|g_{RR}^{S,\ell^{\prime}\ell}|^2+|g_{LL}^{V,\ell^{\prime}\ell}|^2\right).
\end{equation}
For the $\tau\to\mu$ decay, experiments find~\cite{Zyla:2020zbs}
\begin{equation}
    \rho_{\tau\to\mu}^{\text{exp}}=0.763\pm0.020,
\end{equation}
whereas for $\mu\to e$ it is~\cite{Bayes:2011zza,Bueno:2011fq}
\begin{equation}
    \rho_{\mu\to e}^{\text{exp}}=0.74997 \pm 0.00026.
\end{equation}
The deviation of the $\rho$ parameter due to the new physics can be expressed through
\begin{equation}
    \Delta\rho_{\ell\to\ell^{\prime}}\equiv\frac{3}{2}{\rm Re}\left[g_{LL}^{V,\text{SM}}\delta g_{LL}^{V,\ell\ell^{\prime}}\right],
\end{equation}
where $g_{LL}^{V,\text{SM}}=1$ and $\delta g_{LL}^{V,\ell\ell^{\prime}}$ is the new contribution due to the presence of the fakeon doublet. The latter is obtained by renormalizing the four-fermion amplitude in the on-shell renormalization scheme used for the SM, and taking the zeroth order in the expansion of the amplitude in powers of $s/m_W^2$, where $s$ is the center-of-mass energy squared. For the $\tau$ decay, we obtain the bound shown in Fig.\ref{fig:g-2_y_vs_mphi} by requiring $|\Delta\rho_{\tau\to\mu}|<0.020$, while for the $\mu$ decay we use $|\Delta\rho_{\mu\to e}|<0.00026$. The limits we find do not constrain the proposed solution.  

\section*{Summary} 
\label{sec:summary}

We have proposed a new solution to the puzzle of the muon anomalous magnetic moment by modelling the physics beyond the SM with purely virtual degrees of freedom: the fakeons. Considering a new fake scalar doublet interacting sizeably only with muons, electroweak gauge bosons and the Higgs field, we have shown that the new $(g-2)_\mu$ measurement can be explained in a large part of the parameter space. The predictions, together with the most important bounds arising from complementary collider and precision observables, are collected in Fig.~\ref{fig:g-2_y_vs_mphi}. 

Unlike for ordinary particles, our results for $(g-2)_\mu$ are not significantly impaired by the constraints. Fakeon masses at, or below, the GeV scale are not excluded by the precision measurements of the $Z$-boson decay width, because the fakeon quantization prescription precludes the production of these particles. The precision tests of lepton universality in $Z$ and $W$-boson decays, sensitive to deviations at the {per-mille \it} level, also fail to exclude our scenario. The collider measurements of $Z\to 4 \mu$ decays test the solution obtained for a degenerate mass spectrum only in a corner of the parameter space with large values of the Yukawa coupling and masses. We find no constraints on the solutions obtained for a mild hierarchy in the fakeon masses. 

In conclusion, the analyzed framework allows for new effects well below the electroweak scale without contradicting the available experimental results. Our work  motivates further studies of the phenomenology of fake particles in the context of the anomalies in low energy physics observables related to muons.

\section*{Acknowledgments} 

We thank Abdelhak Djouadi, Alessandro Strumia and Hardi Veerm\"ae for useful discussions. This work was supported by the Estonian Research Council grants PRG356, PRG434, PRG803, MOBTT86, MOBTT5 and by the EU through the European Regional Development Fund CoE program TK133 ``The Dark Side of the Universe". 


\bibliography{fakeons}
\bibliographystyle{JHEP}

\end{document}